\documentstyle[a4,12pt]{article}

\newcommand{\smdag}{\mbox {\tiny \dag}}

\begin{document}
\vspace{3cm}
\begin{center}
{\Large{\bf STRONG PIONIC DECAYS OF BARYONS FROM A SPECTROSCOPIC QUARK MODEL}}
\end{center}

\vspace{2cm}

\begin{center}
{\large{F. Cano, P. Gonz\'alez and S. Noguera}}
\end{center}

\begin{center}
{\small{Departamento de F\'{\i}sica Te\'orica and IFIC\\
Centro Mixto Universidad de Valencia - CSIC\\
46100 Burjassot (Valencia), Spain.}}
\end{center}

\vspace{.25cm}
\begin{center}
{\large{B. Desplanques}}
\end{center}

\begin{center}
{\small{Institut des Sciences Nucl\'eaires, F-38026 \\
Grenoble Cedex, France.}}
\end{center}
\vspace{2cm}

\begin{abstract}

	From a refined non--relativistic quark model that fits the baryonic 
low--energy spectrum the study of strong pion 
decay processes within an elementary
emission model scheme points out the need of incorporating size--contributing 
components into the baryon wave functions. In particular the effect of a
($qqq \; q\bar{q}$) component is investigated in the framework of a quark 
pair creation model.
 
\end{abstract}

\vfill
\hfill
 Preprint {\bf FTUV/95-71, IFIC/95-74} 
\newline
\hspace*{\fill} Preprint {\bf arch-ive/9606038}

\newpage

\section{Introduction}

	In QCD, the basic theory of strong interactions (responsible for
the hadronic structure), elementary constituents (quarks) which are massless if
chiral symmetry is preserved can acquire a dynamically generated mass associated
to the spontaneous breakdown of chiral symmetry that takes place when the 
strong interaction reaches a critical intensity. On the other hand, although not
conclusively proved in (3+1) dimensions, QCD seems to be a confining theory. 
Assuming that the energy scale for the chiral symmetry breaking is higher 
than the corresponding one for confinement 
($\simeq \Lambda_{\mbox \scriptsize {QCD}}$), one gets the image of a 
baryon as a confined system of interacting massive quarks and Goldstone
bosons coming from the symmetry breakdown. 

	Little more than this qualitative statement can be said from QCD at
the current moment since the lack of an adequate method to describe a bound
state within a quantum field theory, together with the non--perturbative
character of the interaction below $\Lambda_{\mbox \scriptsize {QCD}}$, makes difficult 
any attempt to go further. As a consequence several models of hadronic structure
have been developed. Among them perhaps the ones closer related to the image 
of QCD just drawn
are the bag models where the quarks move almost freely inside a 
confinement region, the bag, surrounded by a mesonic cloud. However the 
separation of the center of mass motion (indispensable to tackle the problem
of baryon structure) added to the fact that one has to 
simultaneously solve for the movement of the quarks and the bag surface, make 
the model technically complex. To avoid these difficulties one could wonder 
whether it would be possible or not to construct an equivalent (regarding the
low--energy description) non--relativistic model. Undoubtedly the extensions and
refinements from the naive quark model of hadron structure \cite{KOK69} are 
steps in this direction. In this line De R\'ujula {\it et al.} \cite{DERUJULA75}
derived an interquark potential from the non--relativistic reduction of the
one gluon exchange diagram in QCD to be considered together with
the phenomenological confining interaction.
The resulting non--relativistic quark model (NRQM) reproduces the image 
of the quark core of the bag, the mesonic cloud having been obviated. The
absence of mesons together with the non--relativistic treatment poses serious 
doubts about its usefulness unless the incorporation of new 
terms in the potential
and/or the effectiveness of the parameters of the model allows to mitigate at
least partially these shortcomings. As an immediate consequence of this way
of proceeding, the constituent 
quarks loose a direct connection to the quark fields of the QCD lagrangian,
representing effective degrees of freedom. This viewpoint is also supported by
the absence of a 
strong necessity to introduce in the interaction between quarks
some terms issued from the one gluon exchange such as the spin--orbit or 
tensor forces \cite{ISGUR,SILVESTRE86}.

	The first requirement to such an effective model should be a good 
description of the hadronic spectrum (at least at low energies) that allows a
correct assignment of the theoretical states to the experimental ones. This 
can be accomplished with relative success for mesons as well as for baryons by
a ``minimal'' model involving confinement plus part of 
the one gluon exchange (OGE) 
potentials \cite{SILVESTRE86}. Nevertheless an endemic problem 
in the low energy spectrum remains, that is the 
impossibility to correctly predict the masses for the first radial excitations in
the baryonic octet and decuplet. The systematic of this wrong prediction led
some authors  \cite{DESPLANQUES92} to add a phenomenological 
three--quark force getting the correct baryonic spectrum (for 
negative and positive parity states with and without strangeness 
simultaneously) up to an excitation 
energy of 0.7 GeV without need to resort to any specific resonance 
mechanism. This model, that will serve as our starting point now, presents
however some conceptual problems which  we shall consider firstly.

	The second requirement is the correct description of the baryon
properties, what involves the wave functions of the baryons on the one hand, 
the operator relative to the process under discussion on the other hand. 
Having a model that correctly describes the low energy baryonic spectrum, it is
appropriate to compare its predictions to those of other models that do not do 
so well for the baryon masses. This may allow us to determine whether there is some bias to be 
introduced in the comparison to experiments due to a poor prediction of 
baryon masses (though some discrepancy may be due to different physical 
ingredients). This comparison was our main intent originally for looking 
at the strong pion decays of baryons. It however appeared that results were
very much sensitive to the microscopic description of the process. This led us
to examine various aspects of the problem, where relativity is often present
in one way or another. The present work is part of a general program tending 
to extract from the comparison of the models to each other and to the data 
some general features of the ``true'' wave functions that in their  
turn should serve as a guide to further refinements of the models themselves.
 The contents of the presentation are organized hereforth as follows. 
In section 2 we review the quark models we make use of 
and their predictions for static properties. In 
sections 3 and 4 we study strong pion decay processes within two different 
approaches, namely the elementary emission model (EEM) and the quark
pair creation model (QPCM) that we modify in order to correct its energy 
dependence. Finally in section 5 we summarize our main conclusions

\section{The Quark Models}

	The non--relativistic quark model of hadron structure describes the 
hadron (color singlet bound state) in terms of a definite number of components
(constituent quarks $q$, and antiquarks $\bar{q}$) that interact through 
an effective potential. The analysis of the mesonic ($q\bar{q}$) and baryonic
($qqq$) spectrum leads to a minimum two--body potential containing the basic
QCD motivated, confining, coulombic (OGE) and spin--spin (OGE) $qq$ 
interactions, of the form:

\begin{equation}
\label{potencialdebhaduri}
V_{I}= \frac{1}{2} \sum_{i<j} 
		\left( -\frac{\kappa}{r_{ij}} + \frac{r_{ij}}{a^{2}} 
		+ \frac{\kappa_{\sigma}}{m_{i} m_{j}} 
		\frac{\exp{-r_{ij}/r_{0}}}{r_{0}^{2} r_{ij}}
		\vec{\sigma_{i}}\vec{\sigma_{j}} -D \right)
\end{equation}

\noindent 
where $D$ is a constant to fix the origin of the potential, $r_{ij}$ is 
the distance between quarks $i$ and $j$ and $\sigma$ denote the Pauli matrices.
The Yukawa form of the spin--spin term replace the $\delta(\vec{r})$ contact 
interaction of the OGE potential \cite{DERUJULA75} to avoid an unbounded 
spectrum when solving the Schr\"odinger equation \cite{BHADURI80}. 
The mass parameters $m_{i,j}$ are chosen to fit the baryon magnetic moments and
the three parameters $a^{2}$, $\kappa=\kappa_{\sigma}$, $r_{0}$, fitted from
the meson spectrum \cite{SILVESTRE86} (see table 1) provide a good 
description of the baryonic octet and decuplet \cite{SILVESTRE86}, a part of which is 
reproduced in fig. 1. A look at this figure makes clear a general deficiency in
the description: the masses of the positive parity excitations that 
correspond to radial excitations of the totally symmetric spatial components of
the wave function are systematically higher (from 200 to 400 MeV) than the 
experimental ones. Any attempt to correct this situation by a refitting of the 
parameters causes undesirable effects on the rest of the spectrum. Then the 
open question refers to the possibility of solving the problem by means of 
the introduction of some physically founded new term in the potential.

	Attending the mixed symmetric spatial structure of the first negative 
parity excitations it becomes obvious that they will be much less affected by
an interaction tending to group
the three quarks than the spatially symmetric ground states and its 
first radial excitations. Having in mind a genuine three quark interaction 
as the exchange of two sigma mesons at the same point, Desplanques {\it et al.}
\cite{DESPLANQUES92} proposed a three--body potential
 
\begin{equation} 
\label{3des}
V_{II}^{(3)} = \frac{1}{2} \sum_{i \neq j \neq k \neq i}
	\frac{V_{0}}{m_{i}m_{j}m_{k}} 
	\frac{e^{-m_{0} r_{ij}}}{m_{0} r_{ij}} 
	\frac{e^{-m_{0} r_{ik}}}{m_{0} r_{ik}} 
\end{equation}

	The low energy baryonic spectrum can be then well reproduced with a 
'two plus three body' interaction:

\begin{equation}
\label{vsigma}
V_{II}=V_{II}^{(2)} + V_{II}^{(3)}
\end{equation}

\noindent 
where $V_{II}^{(2)}$ differs from $V_{I}$ in the values of the parameters.
As a consequence the unified description of both the mesons and baryonic 
spectra given by $V_{I}$ is lost. More than a drawback one should consider 
this as the natural outcome of the effectiveness of the parameters (through 
them for instance, some mesonic effects, only present in the baryonic case
might be incorporated).

	The form used in ref. \cite{DESPLANQUES92} for the potential, eq. 
(\ref{vsigma}), shows some non--appealing features. Concerning the 
two--body piece $V_{II}^{(2)}$, the range of the spin--spin
interaction, $r_{0}=1.28$  fm, needed in \cite{DESPLANQUES92} is very far 
from the one which could reasonably be expected  
given the original $\delta(\vec{r})$
form of this interaction \cite{DERUJULA75}. This comes from the minimum 
number of parameters' constraint $\kappa=\kappa_{\sigma}$. Nevertheless
one can get a much more ``reasonable'' range (0.4 fm) without increasing the 
number of parameters by recovering the original OGE potential relation between 
the coulombian intensity $\kappa$ ($\kappa = - \frac{4}{3} \alpha_{s}$) 
and the spin--spin one $\kappa_{\sigma}$ 
($\kappa_{\sigma} = - \frac{2}{9} \alpha_{s}$),
i.e. $\kappa_{\sigma}=\frac{1}{6} \kappa$. The resulting value of $r_{0}$
might represent somehow an effective average between 0 fm, the OGE 
$\delta$--term range,
and $\approx$ 1 fm, the range of the  quark pionic exchange not implemented in 
our model. This impression seems to be reinforced by the corresponding value of
the effective QCD coupling constant,
$\alpha_{s} \approx 1$, that doubles the one 
needed in models where pionic quark 
exchanges are explicitly incorporated \cite{VALCARCE}.

	From now on we shall call $V_{II}$ the two plus three body interaction 
with $V_{II}^{(2)}$ corrected as explained. 
The fit to the spectrum with this corrected potential
(see table 1) a part of which is reproduced in fig. 2, hardly differs 
from that in ref \cite{DESPLANQUES92}.

	Regarding the 3--body piece of the interaction
(\ref{vsigma}), $V_{II}^{(3)}$, its
long range, 0.8 fm, can find a justification attending the small value of
the coefficient of the linear potential (or equivalently its small 
contribution to the string tension). For higher angular momentum states 
where the quarks are very far apart, the long range of  $V_{II}^{(3)}$ 
would allow a contribution to the string tension to explain the leading 
Regge trajectories. It is difficult however to justify if we think of the 
three nucleon force this 3--quark term would give rise to. To correct for this 
behaviour a less singular form of the interaction is required. For the sake
of technical simplicity we shall adopt here a gaussian form

\begin{equation} 
\label{3gauss}
V_{III}^{(3)} = 
V_{0}\exp \left( -\sum_{i<j}\frac{r_{ij}^{2}}{\lambda^{2}}\right)
\end{equation}

\noindent 
very easy to be used in the hyperspherical harmonic formalism (Appendix A) 
\cite{BALLOT} we work with.

	The parameters of the potential 

\begin{equation}
\label{vgauss}
V_{III}=V_{III}^{(2)} + V_{III}^{(3)}
\end{equation}

\noindent 
appear in table 1 (note that the range of $V^{(3)}$ has gone down to 0.25 fm
and that we have relaxed the constraint between $\kappa$ and $\kappa_{\sigma}$),
and the comparative spectrum, of similar quality to the $V_{II}$'s one, appears
(in part) in fig. 3.

	With regard to other parameters of the potential some 
comments are in order.
For the quark masses, $m_{u}$, $m_{d}$, a value between 300 and 360 MeV, that
fits the nucleon magnetic moments (assuming an SU(6) spatially symmetric 
wave function) within a 10 \% of error, is generally accepted. It can be
shown (Appendix A) that, from the potential we are dealing with, one can obtain
via an adequate redefinition of the parameters of the potentials, a 
non--strange energy spectrum (for quark mass $m'$) whose energies, $E'$, are related
to the initial ones $E$ (for quark mass $m$) by

\begin{equation}
\label{eqmass}
E'=\frac{m}{m'}E
\end{equation}  

\noindent 
and with no other change than a distance rescaling 
in the wave functions. Hence 
given a fit to the spectrum for a
given quark mass, eq. (\ref{eqmass}) gives the shifted relative energies
when changing the quark mass, i.e. the baryon magnetic moment.

	The strength of the confining term is considerably reduced due to
the presence of the three body force (see table 1). This gives rise to
a grouping
of the states in the 'high' energy part of the spectrum that could be 
suggesting the need of 
some qualitative change in the three--body or in the confinement 
pieces. Nevertheless the lack of precise experimental information in this
region together with the very probable presence of significant relativistic 
effects prevent us from extracting quantitative conclusions from it.

Once, via the fit of the spectrum, some physical states are unambiguously 
ascribed to model states (up to 0.7 GeV  excitation energy)
we should focus on the description of the 
baryonic properties. As explained before the nucleon magnetic moments are 
reasonably well reproduced (see table 2) with masses between 300 and 
360 MeV indicating the dominance of the spatially symmetric component of 
the wave function (see fig. 4). A different case is the mass and charge 
root mean square radius
(r.m.s.). For the nucleon charge radius for instance the predicted values
are too low compared to the data. 
This can be understood by taking into account that our 
effective scheme only incorporates the description of the core of quarks,
whose pretty small size has to do with the pressure exerted by the mesonic
cloud that in our meson--absent models is simulated through the effective 
values of the parameters. Then, 
if the mesonic contribution to the total mass is 
small as compared to the core contribution (as it happens to be the case in 
the bag models), one can understand the r.m.s.--spectrum puzzle, i.e. the 
impossibility within a model scheme that only plays with effective quark
degrees of freedom of simultaneously fitting the spectrum and the total sizes.

	Furthermore the absence of ($qqq \; q\bar{q}$) components in the
wave function should be at least in part also responsible for the high values 
(5/3) predicted for the axial coupling constant $g_{A}$ as it is suggested by
bag model calculations \cite{VENTO83}.

	As a corollary it may be established that when dealing with 
physical processes where the mesonic cloud plays a relevant role as such 
the use of a spectroscopic quark model
requires the use of effective operators that can give account of
it. Hence for the strong pion decay we consider next a careful analysis of the 
transition matrix elements is needed before a comparison between different 
spectroscopic quark models makes any sense.

\section{The Elementary Emission Model (EEM)}  

	To study strong baryonic decays we shall follow the elementary emission
model approach developed long time ago \cite{FAIMAN69}. The decay takes place
through the emission of a point--like pion by one of the quarks of the baryon 
(fig. 5) and therefore the coupling constant is unique for all the processes.

	Up to order ($\frac{p_{q}}{m_{q}}$) the matrix elements for the process
$B \rightarrow B'\pi$ can be expressed as $\langle B' | H | B \rangle$ 
\cite{LEYAOUANC88} with 

\begin{eqnarray}
\label{notengomaslabeles}
H & = &
- \frac{3i}{(2\pi)^{3/2}} \frac{1}{(2 \omega_{\pi})^{1/2}} 
\frac{f_{qq\pi}}{m_{\pi}}(\tau^{\alpha\; (3)})^{\smdag} \nonumber \\
&  &  \left[ \vec{\sigma}^{(3)} \vec{k} 
e^{-i \vec{k} \vec{r}_{3}}
- \frac{\omega_{\pi}}{2 m_{q}}\left\{ \vec{\sigma}^{\;(3)}\vec{p}^{\;(3)},
 e^{-i \vec{k} \vec{r_{3}}} \right\} \right]
\end{eqnarray}

\noindent 
where the factor 3 comes from the number of quarks,
$\omega_{\pi}$ and $\vec{k}$ stand for the pion energy and trimomentum
respectively and $\vec{r}_{3}$ is the quark 3 coordinate.
The upper index (3) on the operators of spin, $\vec{\sigma}$, isospin, 
$\tau$, and quark trimomentum, $\vec{p}$, indicates
that we have chosen for later technical simplicity the quark 3 to perform
the calculations. The coupling constant $f_{qq\pi}$ is related to 
the usual $g_{qq\pi}$ through the quark--pion mass ratio:

\begin{equation}
\frac{f_{qq\pi}}{m_{\pi}} = \frac{g_{qq\pi}}{2 m_{q}}
\end{equation}

\noindent 
and is the only free parameter of the model. 

	The `non--relativistic' $qq\pi$ interaction given by $H$ 
may be obtained from the $qq\pi$ invariant interaction
with the pseudovector coupling 
${\cal L}_{qq\pi} \simeq \bar{\Psi}_{q}
\vec{\tau} \gamma_{\mu} \gamma_{5} \Psi_{q} \partial^{\mu} \vec{\pi}$. It 
differs from that obtained with the pseudoscalar coupling  
${\cal L}_{qq\pi} \simeq \bar{\Psi}_{q}
\vec{\tau} \gamma_{5} \Psi_{q} \vec{\pi}$ since at first order in 
$(\frac{p}{m})$, this one does not give rise to the term proportional to 
$\omega_{\pi}$ in eq. (\ref{notengomaslabeles}) (it only gives the 
$\vec{\sigma} \vec{k}$ term). The examination of the higher order terms in 
$(\frac{p}{m})$ in the pseudoscalar coupling case shows that there are some 
$(\frac{p}{m})^{2}$ corrections to the $\vec{\sigma} \vec{k}$ term which are 
identical to those  that would be derived 
in the pseudovector coupling. As an expansion in 
$(\frac{p}{m})^{2}$ is meaningless in view of the large value taken by this
quantity (larger than 1, with often a destructive interference at the first 
order) we prefer to freeze these $p^{2}$ terms and replace them by some 
constant. In practice, this means that they will be hidden in the coupling 
$f_{qq\pi}$ to be fitted to reproduce the $f_{NN\pi}$ coupling.
The other $(\frac{p}{m})^{2}$ corrections in the pseudoscalar coupling 
represent the difference in the
kinetic energies of quarks in the initial and final states, which enters in
$\omega_{\pi}$ ($\omega_{\pi} = E_{B} - E_{B'}$). The difference in potential
energies in the initial and final states, which is also included in 
$\omega_{\pi}$, is provided by terms involving the excitation of a 
quark--antiquark pair (this one is known to be important in the 
pseudoscalar coupling). 

	Notwithstanding that it corresponds to a higher order term in 
$(\frac{p}{m})$ in the pseudoscalar coupling, we
shall maintain the term proportional to $\omega_{\pi}$ in eq. 
(\ref{notengomaslabeles}), because of its different functional structure.
Altogether, eq.   (\ref{notengomaslabeles}) amounts to retain for the different
types of terms, $\vec{\sigma} \vec{k}$ and $\vec{\sigma} \vec{p} \;$, 
corresponding to the 
lowest order contributions in $(\frac{p}{m})$, relativistic corrections being 
embedded in the parameter $f_{qq\pi}$ or in the factor $\omega_{\pi}$ itself. 

	In order to evaluate the matrix elements of the hamiltonian $H$ the 
wave function of a baryon of spin $J$ and third component $J_{z}$ is factorized
as a plane wave of trimomentum $\vec{P}$ (the baryon trimomentum) for the 
center of mass motion multiplied by the intrinsic wave function, i.e.

\begin{equation} \label{wavegeneral}
|B \rangle = \frac{1}{(2 \pi)^{3/2}} e^{i\vec{P}_{B}\vec{R}}
\left[ \Psi_{B}(\vec{\xi}_{1},\vec{\xi}_{2}) \Phi_{B} (S,M_{S};I,M_{I}) \right]
_{JJ_{z}}
\end{equation}

\noindent 
where $\Phi_{B}$ stands for the spin--isospin part and $\Psi_{B}$ is the 
spatial internal wave function in terms of the Jacobi coordinates 
$\vec{\xi}_{1}$, $\vec{\xi}_{2}$.

	Then the amplitudes for the decay, defined by:

\begin{equation}
\label{helicidad}
_{J',\lambda} \langle B' | H | B \rangle_{J,\lambda} =
\frac{1}{(2 \pi)^{3/2}} \delta(\vec{P}_{B} - \vec{P}_{B'} - \vec{k})
A_{\lambda}^{(i,f)}
\end{equation}

\noindent 
where $i$ and $f$ refer to the internal quantum numbers of the initial and 
final baryons, are given by:

\begin{eqnarray}
\label{amplitudgorda}
A_{\lambda}^{(i,f)} & = & - \frac{3 i}{(2 \omega_{\pi})^{1/2}} \frac{f_{qq\pi}}{m_{\pi}} 
\int d\vec{\xi}_{1} \; \int d\vec{\xi}_{2} \; 
\left[ \Psi_{B'}(\vec{\xi}_{1},\vec{\xi}_{2}) \Phi_{B'} 
(M',M_{S}';I',M_{I}') \right]_{J'\lambda}^{*} \nonumber \\ 
&  &\;\;\;\;\;\;\;(\tau^{\alpha\; (3)})^{\smdag} 
\left[ \vec{\sigma}^{(3)} 
\vec{k} \left( 1 + \frac{\omega_{\pi}}{6 m_{q}}\right) 
e^{i \sqrt{\frac{2}{3}}\vec{k} \vec{\xi}_{2}} \nonumber \right. \\
& &  \left. \;\;\;\;\;\;\;\;\;\;\;\;\;\;\;\;\;\;\;\;\;\; - \frac{\omega_{\pi}}{2 m_{q}}\left\{ \vec{\sigma}^{\;(3)}
\left( -\sqrt{\frac{2}{3}} \vec{p}_{\xi_{2}} + \frac{\vec{P}_{B}}{3}  \right),
 e^{i \sqrt{\frac{2}{3}} \vec{k} \vec{\xi_{2}}} \right\} \right] \cdot \nonumber \\
&  &   \;\;\;\;\;\;\; \left[ \Psi_{B}(\vec{\xi}_{1},\vec{\xi}_{2}) \Phi_{B} 
(S,M_{S};I,M_{I}) \right]_{J\lambda} 
\end{eqnarray}

	The dependence on $\vec{P}_{B}$ may suggest a dependence of the amplitudes
on the frame of reference. In the present case, the corresponding term combines
with the first term proportional to $\vec{k}$ in eq. (\ref{amplitudgorda}) to get
a term proportional to $\vec{k} - \frac{\omega_{\pi}}{3 m_{q}} \vec{P}_{B}$ 
($\simeq \vec{k} - m_{\pi} \frac{\vec{P}_{B}}{M_{B}}$) in the non relativistic 
limit), which ensures the Galilean invariance of the amplitude at the lowest 
order in $\frac{\omega_{\pi}}{M_{B}}$. This invariance of the amplitude at the 
quark level was imposed by Mitra and Ross \cite{MITRA} to introduce in the 
$qq\pi$ interaction some $\omega_{\pi}$ dependent term, quite similar to the
one in eq. (\ref{notengomaslabeles}). We shall work in the frame
where the decaying resonance
is at rest ($\vec{P}_{B} = \vec{0}$). For later purpose it is convenient to reexpress the
amplitudes in the more compact form:

\begin{equation}
\label{parametrizacion}
A_{\lambda} = - \frac{3 i}{(2 \omega_{\pi})^{1/2}}
\frac{f_{qq\pi}}{m_{\pi}} 
\left\{\left( 
1 + \frac{\omega_{\pi}}{6 m_{q}}\right) N_{i,f} -
\frac{\omega_{\pi}}{2 m_{q}}  R_{i,f} \right\}
\end{equation}

The notation $N_{i,f}$, $R_{i,f}$ 
is a technical one. $R$ includes derivatives (coming from the momentum 
operator $\vec{p}_{\xi_{2}}$) whereas $N$ does not. They are directly related
to the more physical recoil (depending on the recoil momentum ($\vec{p}^{(3)}-
\vec{k}$)) and direct (depending of the pion momentum $\vec{k}$) contributions.

	Finally the decay width is obtained from the amplitudes in the initial 
baryon (B) rest frame as \cite{LEYAOUANC88}:

\begin{equation}
\Gamma_{B \rightarrow B' \pi} = \frac{1}{ 2 \pi} 
\frac{E_{B'} \omega_{\pi}}{m_{B}} k \overline{\sum_{\lambda,i,f}} 
|A_{\lambda}^{(i.f)}|^{2}
\end{equation}

\noindent 
where the overline indicates the average over the initial spin--isospin. The 
kinematical magnitudes and the baryon masses depend on the spectroscopic 
model, so that for $V_{II}$ and $V_{III}$ they are pretty close to the 
experimental ones, whereas for $V_{I}$ the difference (especially for the
Roper decays) can be very significant. 
Nevertheless as we shall be interested (once the transition operator has
been chosen) in testing the baryonic wave 
functions we shall use the same kinematics, i.e. the experimental values of 
$m_{B}$, $E_{B'}$, $\omega_{\pi}$ and $\vec{k}$, in all our calculations. 

\subsection{Fitting of $f_{qq\pi}$}

	There is no unique criterium in the literature to fix $f_{qq\pi}$. 
For instance $f_{qq\pi}$ could be chosen to get the best global fit to the 
data \cite{KONIUK80}. Nonetheless, for the purpose of analyzing the 
baryonic wave function another possibility is more convenient, namely to 
fit $f_{qq\pi}$ to reproduce the $NN\pi$ interaction at low momentum transfers.
Thus we compare the matrix element for the process 
$p \uparrow \rightarrow p \uparrow \pi^{0}$ calculated
at the quark level from H, eq. (\ref{notengomaslabeles}), with the same matrix
element at baryonic level obtained from:

\begin{eqnarray}
\label{hnnpi}
H_{\mbox {\scriptsize Bar}} & = & - \frac{i}{(2 \pi)^{3/2}}
\frac{1}{(2 \omega_{\pi})^{1/2}} \frac{f_{NN\pi}(k)}{m_{\pi}}
(\tau_{N}^{\alpha})^{\smdag} \nonumber \\
 &  & \left[  (\vec{\sigma_{N}}\vec{k}) e^{-i \vec{k}\vec{R}}
- \frac{\omega_{\pi}}{2 m_{N}} \left\{    
\vec{\sigma}_{N}\vec{P}_{N} , 
e^{-i \vec{k}\vec{R}} \right\}     \right]
\end{eqnarray}

	By assuming the nucleon to be a totally symmetric spin--isospin
SU(6) state (the mixed--symmetric component of the nucleon has less than 2 \%
probability for our quark models) one gets:

\begin{equation} 
\label{mystrongff}
f_{NN\pi}(k) = f_{qq\pi} \frac{5}{3} 
\frac{\left(1 + \frac{\omega_{\pi}}{6 m_{q}}\right)}
{\left(1 + \frac{\omega_{\pi}}{2 m_{N}}\right)} F(k)
\end{equation}

\noindent 
where $F(k=|\vec{k}|)$ 
contains all the information about the spatial structure
of the nucleon as provided by the quark model. Explicitly:

\begin{equation}
\label{estructuraradial}
F(k)= \frac{12}{k^{2}} {\cal I}_{N_{1}N_{1}}^{3,2}(k)
\end{equation}

\noindent
where $\cal I$ represents an integral of the general form:

\begin{equation}
{\cal I}_{B_{i} C_{j}}^{l,m} (k) = \int d\xi \; \xi^{l} \Psi_{B_{i}}^{*}(\xi)
J_{m}(\sqrt{\frac{2}{3}} k \xi) \Psi_{C_{j}}(\xi)
\end{equation}

\noindent 
$J_{m}$ standing for the Bessel function of order m and $\Psi_{B_{i}}$
($\Psi_{C_{j}}$) for the hyperradial part of the wave function for the 
$i$ ($j$) channel of baryon B (C).

	In the limit ($\vec{k} \rightarrow 0$, $\omega_{\pi} = m_{\pi}$), 
the spatial structure reduces to $F(k) \rightarrow 1$, and then

\begin{equation}
f_{qq\pi} = \frac{3}{5} \frac{f_{NN\pi}(0)}{B(m_{\pi})}
\end{equation}

\noindent 
having defined 

\begin{equation}
B(x) = \frac{(1 + \frac{x}{6 m_{q}})}{(1 + \frac{x}{2 m_{N}})}
\end{equation}

	Hence the only dependence of $f_{qq\pi}$ on the spectroscopic
model comes from the non--relevant quark mass differences, having the values
$f_{qq\pi}= 0.602, 0.604, 0.600 $ for $V_{I}$, $V_{II}$ and
$V_{III}$ respectively ($f_{NN\pi}=0.998$). 

	Once $f_{qq\pi}$ has been fitted in this almost model 
independent manner, the predictions of the models serve as 
a test, within a EEM scheme, of the 
baryonic wave functions. For instance, by writing $f_{NN\pi}$ in terms of a 
normalized form factor $G(k)$:

\begin{equation}
\label{ffnor}
f_{NN\pi}(k) = f_{NN\pi}(0) G(k)
\end{equation}

\noindent 
with $G(k=0,\; \omega_{\pi}= m_{\pi}) =1$, we have

\begin{equation}
G(k)=F(k)\frac{B(\omega_{\pi})}{B(m_{\pi})}
\end{equation}
  
	The results appear in fig. 6, against the phenomenological 
parameterization of ref. \cite{OSET83}. As can be seen neither $V_{II}$ nor
$V_{III}$ provide with a good form factor (although the deviation is at
most of a 10 \% for small k values). The reason for it is clear when making
an expansion in powers of k:

\begin{equation}
G(k) = 1 - \frac{1}{6} k^{2} \langle r^2 \rangle + \ldots 
\end{equation}

	Thus for the small values of k one is directly testing the root mean
square mass radius:

\begin{eqnarray}
\langle r^2 \rangle = \langle
\frac{\sum_{i=1}^{3} m_{i} (\vec{r}_{i} - \vec{R})^{2}}
{\sum_{i=1}^{3} m_{i}}  \rangle & = & \frac{1}{3} \langle \xi^2 \rangle 
\label{na}
\end{eqnarray}

\noindent
which turns out to be too small for $V_{II}$ and $V_{III}$. 
This is not certainly a surprise from our 
considerations in section 2 since it can be an indication that relativistic
effects, pionic components ($qqq \; q\bar{q}$), etc may be playing 
a role. Alternatively it may be suggesting the need to refine the three--body
potential.

\subsection{Strong pion decays} 

	To pursue the analysis of the wave functions, within the EEM, we evaluate
the widths for strong pion decays of nucleon and delta resonances. The results 
are
listed in table 3. Although it is difficult to extract general features
from them (the almost zero widths for the processes involving the ground state 
and its first radial excitation, $N(1440) \rightarrow N\pi$ and 
$\Delta(1600) \rightarrow \Delta\pi$, comes from the orthogonality 
of their radial wave functions) we can say that for decays clearly dominated 
by the non--derivative part of the amplitude ($N(1520) \rightarrow N\pi$, 
$\Delta(1232) \rightarrow N\pi$) the three 
potentials work reasonably well, the better the result the bigger the predicted
radius for the nucleon. This comes along the same line that our previous 
discussion for the $NN\pi$ form factor since the non--derivative terms involve 
the average wave function overlap. In order to try to include size contributing
components, we shall explore next the possibility of 
incorporating the pion structure in a non--relativistic scheme.

\section{The Quark Pair Creation Model (QPCM)}

	To implement the mesonic structure several formalisms have been
developed \cite{LEYAOUANC88}, all of them sharing the image of the meson 
emission as the creation of a $q\bar{q}$ pair that by later recombination
gives rise to the outgoing meson. 

	At the effective level we work, the choice of one or other 
QPCM comes mainly motivated by simplicity. In this sense the $^{3}P_{0}$ 
QPCM is manageable and comparable in the limit of point pion to the EEM.
Schematically the process is pictured in fig. 7. The $q\bar{q}$
pair created has the quantum numbers of the vacuum: flavor and color 
singlet, zero momentum and total angular momentum $J^{PC}=0^{++}$ 
($\Rightarrow L=1, S=1$). This translates into a transition operator:

\begin{eqnarray}
T & = & -\sum_{i,j} \int \; d\vec{p}_{q}  d\vec{p}_{\bar{q}} \; \left[
3 \gamma \delta(\vec{p}_{q} + \vec{p}_{\bar{q}}) \sum_{m}
(110|m,-m) {\cal Y}_{1}^{m}(\vec{p}_{q} -\vec{p}_{\bar{q}})
{\cal Z}_{i,j}^{-m} \right] \cdot \nonumber \\ 
& & \;\;\;\;\; b^{\smdag}_{i}(\vec{p}_{q})
d^{\smdag}_{j}(\vec{p}_{\bar{q}})
\end{eqnarray}	

\noindent
where $\gamma$ is the (dimensionless) parameter strength of the model,
${\cal Y}_{L}^{M}$ is a solid harmonic and ${\cal Z}_{i,j}^{-m}$ contains
the color--spin--isospin wave function of the pair. The matrix element for the
process $B \rightarrow B' M$ is then written as:

\begin{equation}
\label{3p0matrix}
\langle B' M | T | B \rangle = - 3 \gamma \sum_{m} (1 1 0| m, -m)
\langle \Phi_{B'} \Phi_{M} | \Phi_{B} \Phi^{-m}_{\mbox {\tiny pair}} \rangle
I_{m}(B; B' M)
\end{equation}  

\noindent
where $\Phi$ stands for the spin--isospin wave function and

\begin{eqnarray}
I_{m} & = & \int d\vec{p}_{1} \; d\vec{p}_{2}  \; d\vec{p}_{3}
 \; d\vec{p}_{4} \; d\vec{p}_{5} \cal{Y}_{1}^{m}(\vec{p}_{4} - \vec{p}_{5})
\delta(\vec{p}_{4} + \vec{p}_{5}) \cdot \nonumber \\
& & \;\;\;\;\;\;\;\;
\Psi_{B'}^{*}(\vec{p}_{1},\vec{p}_{2},\vec{p}_{4})
\Psi_{M}^{*}(\vec{p}_{3},\vec{p}_{5})
\Psi_{B}(\vec{p}_{1},\vec{p}_{2},\vec{p}_{3})
\nonumber \\ 
& = & \frac{1}{(2 \pi)^{3/2}} \delta(\vec{P}_{B} - \vec{P}_{B'} - 
\vec{k}) \int d\vec{\xi}_{1} \; d\vec{\xi}_{2} \; d\vec{\xi}^{\; '}_{2}
\Psi_{B'}^{*}(\vec{\xi}_{1} ,\vec{\xi}_{2}^{\; '}) 
{\sl O}(\vec{\xi}_{2}^{\; '}, \vec{\xi}_{2}) 
\Psi_{B}(\vec{\xi}_{1} ,\vec{\xi}_{2}) 
\end{eqnarray}

\noindent
the kernel ${\sl O}(\vec{\xi}_{2}^{\; '}, \vec{\xi}_{2})$ being the 
non--local operator:

\begin{eqnarray}
{\sl O}(\vec{\xi}_{2}^{\; '}, \vec{\xi}_{2}) & = & 
 3 {\cal Y}_{1}^{m}(- \sqrt{\frac{2}{3}} (\vec{p}_{\xi_{2}} +
\vec{p}_{\xi_{2}'} )
 - \frac{4}{3} \vec{k} + \frac{2}{3} \vec{P}_{B})
\;\Psi_{M}^{*}(\sqrt{\frac{3}{2}} (\vec{\xi}_{2}^{\; '} - \vec{\xi}_{2}))
\nonumber \\
& &  \hspace{-1cm} 
\exp\left(i \sqrt{\frac{3}{2}} \vec{P}_{B} (\vec{\xi}_{2}^{\; '} - \vec{\xi}_{2}) \right)
\exp\left(i \sqrt{\frac{1}{24}} \vec{k} (\vec{\xi}_{2}^{\; '} - \vec{\xi}_{2})\right)
\exp\left(i \sqrt{\frac{2}{3}} \vec{k} \vec{\xi}_{2}\right)
\label{non--local}
\end{eqnarray} 

\noindent
in terms of the meson wave function $\Psi_{M}$.

	Let us note that,
in the rest frame of the decaying
baryon ($\vec{P}_{B}=0$) we recover in the limit of a point--like meson 
$\Psi_{M}(\sqrt{\frac{3}{2}} (\vec{\xi}_{2}^{\; '} - \vec{\xi}_{2}))
\rightarrow (2 \pi)^{3/2} \sqrt{\frac{2}{3}} 
\delta(\vec{\xi}_{2}^{\; '} - \vec{\xi}_{2})$
the operational structure of the EEM 
(see eqs. \ref{helicidad} and \ref{amplitudgorda}), say

\begin{eqnarray}
\langle B' M | T | B \rangle & \rightarrow &
\frac{1}{(2 \pi)^{3/2}} \delta(\vec{P}_{B} - \vec{P}_{B'} - 
\vec{k}) (- \gamma 3 \sqrt{3} \pi)  
\langle \Phi_{B'} \Phi_{M} | \Phi_{B} \vec{\Phi}_{\mbox {\tiny pair}} \rangle 
\nonumber \\
& & \hspace{-2cm} \int d\vec{\xi}_{1} \; d\vec{\xi}_{2} 
\Psi_{B'}^{*}(\vec{\xi}_{1} ,\vec{\xi}_{2}) 
\left[  \frac{4}{3} \vec{k} e^{i \sqrt{\frac{2}{3}}\vec{k}\vec{\xi}_{2}}
+ \sqrt{\frac{2}{3}} \left\{    
\vec{p}_{\xi_{2}} , e^{i \sqrt{\frac{2}{3}}\vec{k}\vec{\xi}_{2}} \right\}
\right] 
\Psi_{B}(\vec{\xi}_{1} ,\vec{\xi}_{2})
\label{ultimahora}
\end{eqnarray}  

	Hence it is obvious that we reproduce the EEM amplitudes by making in 
eq. (\ref{amplitudgorda}) the replacements:

\begin{eqnarray}
\frac{\omega_{\pi}}{2 m_{q}} & \longrightarrow & 1 \\
\label{replacements}
\frac{3 i f_{qq\pi}}{(2 \omega_{\pi})^{1/2} m_{\pi}} 
& \longrightarrow & \gamma 3 \sqrt{3} \;
\frac{\pi}{4} 
\end{eqnarray}
 
	In other words the compositeness of the meson reflects, with regard
to the EEM, not only through the meson wave function but also in the 
transition ope\-ra\-tor. 

	For the pion wave function we shall make use of a 
gaussian form fitted ($R_{A} = 8$ GeV$^{-2}$) to reproduce the 
root mean square radius of the pion \cite{LEYAOUANC73}:

\begin{equation}
\Psi_{\pi}(\vec{r}) = \frac{1}{(\pi R_{A}^{2})^{3/4}} 
\exp \left( -\frac{\vec{r}^{\; 2}}{2 R_{A}^{2}}\right)
\label{pgaussian}
\end{equation}

	With an orientative purpose we compare it with the wave function 
extracted from the knowledge of the
electromagnetic pion from factor for which (at low $Q^2$) we can very 
approximately assume vector meson ($\rho$--meson) dominance \cite{BHADURI86}:

\begin{equation}
F_{\pi}(Q^{2}) = \frac{m_{\rho}^{2}}{m_{\rho}^{2} + Q^{2}}
\label{electroff}
\end{equation}

	The Fourier transform of $F_{\pi}(Q^{2})$ gives the charge density
from which the pion wave function can be derived:

\begin{equation}
\Psi_{\pi}(\vec{r}) = Y_{00}(\hat{r}) \frac{m_{\rho}}{r^{1/2}} 
\exp \left( -\frac{m_{\rho} r}{2} \right)
\label{pvmd}
\end{equation}
 
	We should keep in mind however that this expression is meaningful 
only at long distances 

	In fig. 8 we draw the two normalized pion wave functions where the 
very different short--distance behaviour can be appreciated.

\subsection{The normalized $NN\pi$ form factor}

	To get the normalized form factor defined previously 
in eq. (\ref{ffnor}), we
compare the matrix element for $p \uparrow \rightarrow p \uparrow \pi^{0}$ 
calculated from $H_{\mbox {\scriptsize Bar}}$ 
(eq. \ref{hnnpi}) with the one provided by the QPCM,  
(eq. \ref{3p0matrix}). The results appear in fig. 9. According to
our expectations the inclusion of the pion structure seems to play the same 
role that the increasing of the baryonic size in the EEM model. 
In this sense half of the difference between the theoretical and the 
phenomenological results has been corrected for $V_{II}$ and $V_{III}$ (for
$V_{I}$ the quantitative change is not very relevant).

\subsection{Fitting of $\gamma$}
 
	Although the formal replacement in (\ref{replacements}) establishes
a relationship between $f_{qq\pi}$ and $\gamma$ it should be handled with care
since it is only valid in the point--like limit. If we
extract $\gamma$ from $f_{NN\pi}(0)$ in the same way as we did to get 
$f_{qq\pi}$ we have:

\begin{equation}
\frac{i}{(2 m_{\pi})^{1/2}} \frac{f_{NN\pi}(0)}{m_{\pi}} \left(
1 + \frac{m_{\pi}}{2 m_{N}} \right) = \frac{15}{12 \pi^{3}} \gamma I(0)
\end{equation} 

\noindent
where
\begin{eqnarray}
I(0) & = & \int d\xi \; d\xi' \; d\phi \; d\hat{\xi}_{2} \; d\hat{\xi}_{2}'
\cos^2\!\phi \; \sin^{2}\!\phi \; \sqrt{\xi^{' \;2} - \xi^{2} \sin^{2}\!\phi}
\; \xi^{5} \xi' \; \Theta\!\left(  1 - \frac{\xi \sin\! \phi}{\xi'}\right) \cdot 
\nonumber \\
 & & \hspace{-1.5cm}\Psi_{\pi}(a) \left[ - \frac{2}{3} \Psi_{N_{1}}(\xi)\Psi_{N_{1}}(\xi') 
 - \frac{1}{12} \left( 3 \xi \cos\!\theta_{2} \cos\!\phi + \xi' 
\cos\!\theta_{2}' 
\sqrt{1 - \left(\frac{\xi \sin\!\phi}{\xi'}\right)^{2}}\right) \cdot
\right. \nonumber \\
 &  &  
\left. \left(  \cos\!\theta_{2} \cos\! \phi \Psi_{N_{1}} '(\xi) \Psi_{N_{1}}(\xi') 
 - \cos\!\theta_{2}'\sqrt{1 - \left(\frac{\xi \sin\!\phi}{\xi'}\right)^{2}} 
\Psi_{N_{1}}'(\xi')\Psi_{N_{1}}(\xi) \right)\right] \nonumber \\ 
\end{eqnarray}

\noindent
$\Theta$ is the Heaviside step function and the argument of the pion 
wave function is 

\begin{equation}
a = \xi^{2} \cos \! 2  \phi + \xi^{'\; 2} - 2 \xi \cos \! \phi
 \sqrt{\xi^{' \;2} - \xi^{2} \sin^{2}\!\phi} \; (\hat{\xi}_{2} \hat{\xi}_{2}')
\end{equation}

	This gives for $\gamma$ the values shown in table 4.
We observe for $V_{I}$ a quite different value of $\gamma$ 
than for $V_{II}$ and $V_{III}$. This comes about due to the introduction
of the pion structure; in the EEM model the constancy of $f_{qq\pi}$ had to do
with the fact that the baryonic quark structure contributes
through model independent normalization integrals whereas in the QPCM case the
presence of the pion wave function weights in a different form the quark
structures giving provoking a model dependence in $\gamma$ 
that reflects the very 
different quark--model baryon wave functions, with and without three--quark
force. 
Concerning the pion wave function its very short distance behaviour 
seems to affect the results very little. It turns out that this will also be
the case for strong pion decays. For this reason we shall restrict hereforth 
our presentation to the gaussian wave function eq. (\ref{pgaussian}). 
Let us notice that the values of 
$\gamma$ available in the literature are much lower than ours, the difference
coming from the quark model used (harmonic oscillator) and from having fitted
$\gamma$ in a different way. In ref. \cite{LEYAOUANC73} the vertex $NN\pi$
was calculated analytically with the harmonic oscillator model. By choosing
a value of $R_{A}^{2}= 8 $ GeV$^{-2}$ for the pion wave function and a nucleon
r.m.s. $R_{N}^{2}= 6 $ GeV$^{-2}$ (very close to the one obtained with
$V_{I}$) the constant $\gamma$ would be 6.29 (6.97 with $V_{I}$).

\subsection{Strong Pion Decays}

	The widths obtained with the $^{3}P_{0}$ model appear in table 
5. We see that the introduction of the pion 
structure does not seem to represent any improvement of the results but 
rather on the contrary a worsening of the fit.

	As the specific form of the pion wave function is not very relevant
we center our attention in the transition operator. Possible corrections to it
may have a relativistic origin and can be enforced on the basis of general principles. First there is a normalization factor 
$\sqrt{\frac{m_{\pi}}{\omega_{\pi}}}$ which is related to the boost of the 
pion from its rest frame to the rest frame of the initial baryon.
On the other hand, if one desires the quark momentum term to have the
same properties as the $\vec{k}$ dependent one under exchanging the
role of initial and final states it has to be associated to the difference
of the energies of the initial and final states, which is nothing but
$\omega_{\pi}$. In the non--relativistic limit, this factor tends to 
$2 m_{q}$ and it is therefore natural to introduce the factor 
$\frac{\omega_{\pi}}{2 m_{q}}$ in front of the momentum term.

	This considerations can be implemented in eq. (\ref{non--local}) by
introducing a factor $\sqrt{\frac{m_{\pi}}{\omega_{\pi}}}$ and 
by multiplying the momentum 
($\vec{p}_{\xi_{2}}+ \vec{p}_{\xi_{2}'} $) dependent term by 
$\frac{\omega_{\pi}}{2 m_{q}}$. Actually by making the additional replacement
of the $\vec{k}$ coefficient
$\frac{4}{3}$ by $\left( 1 + \frac{\omega_{\pi}}{6 m_{q}}\right)$ these
changes are equivalent to modify the QPCM to get as the point pion limit the 
EEM. Adopting this point of view 
we define our $^{3}P_{0}$ modified QPCM  through the operator:

\begin{eqnarray}
{\sl O}_{\mbox {\tiny mod}}(\vec{\xi}_{2}^{\; '}, \vec{\xi}_{2}) & = & 
 3 \sqrt{\frac{m_{\pi}}{\omega_{\pi}}} 
{\cal Y}_{1}^{m}(- \frac{\omega_{\pi}}{2 m_{q}}\sqrt{\frac{2}{3}} 
(\vec{p}_{\xi_{2}} + \vec{p}_{\xi_{2}'})
 - \left( 1 + \frac{\omega_{\pi}}{6 m_{q}}\right)
\vec{k} + \frac{2}{3} \vec{P}_{B}) \cdot
\nonumber \\
& & 
\Psi_{M}^{*}(\sqrt{\frac{3}{2}} (\vec{\xi}_{2}^{\; '} - \vec{\xi}_{2}))
\exp\left(i \sqrt{\frac{3}{2}} \vec{P}_{B} (\vec{\xi}_{2}^{\; '} - \vec{\xi}_{2}) \right) 
\cdot \nonumber \\
 & & \exp\left(i \sqrt{\frac{1}{24}} \vec{k} (\vec{\xi}_{2}^{\; '} - \vec{\xi}_{2})\right)
\exp\left(i \sqrt{\frac{2}{3}} \vec{k} \vec{\xi}_{2}\right)
\label{non--local2}
\end{eqnarray} 

	By repeating the calculational process as explained in section 4.1 and
4.2 the improvement of the fit with this
modified QPCM is spectacular as seen in table 6 (the normalized form factor 
hardly changes and we do not draw it again). Except for 
$\Delta(1600) \rightarrow N\pi$ and $N(1535) \rightarrow \Delta\pi$ where 
the strong cancellation between derivative and non--derivative terms makes
the result very sensitive to their precise values all the widths are
reproduced withing a factor 2. (Let us remark that
the experimental interval has been taken as the most--restrictive one, i.e.
by applying the corresponding variable decay percentage to the central value
of the width). 

	The results for the transitions involving the nucleon (or $\Delta$ 
particle) and its radial excitations, $N(1440)$ (or $\Delta(1600)$) should
be noticed. The finite size of the pion makes them sizeable and in rough
agreement with experiment. This is achieved without relying on 
$(\frac{p}{m_{q}})^{2}$ corrections which in the EEM would probably allow
one to get a similar result. Indeed, the orthogonality of wave functions,
which explains the low transition rates given in table 3, does not 
apply anymore when the operator nature of these corrections is accounted
for.

	The overall results for the three potentials do not differ very much. 
Added to the $NN\pi$ form factor prediction and keeping in mind that 
other corrections, as the ones coming from the
$\Delta$--width in the final states, could be evaluated, the small
remaining discrepancies with experimental data might then have to do with a
fine tuning improvement of the quark potential on the base of increasing the 
quark model radii.

\section{Summary}

	A refinement of a non--relativistic quark model that by means of the 
incorporation of a three quark potential is able to reproduce the nucleon and
delta low energy spectra has been carried out avoiding some 
conceptual difficulties associated to the very singular form of the three--body
force and to the values of the potential parameters in the original version.
The predictions for static properties point out the need of incorporating 
($qqq \;q\bar{q}$) components into the wave function. An effective manner to do
this when studying strong decay processes is through the transition operator 
defining the calculation scheme either via a modification of the coupling
constant as in a elementary emission model (EEM)
or additionally via the explicit implementation of $q\bar{q}$ 
pair creation as in the quark pair creation models (QPCM).

	We do find some sensitivity to the way the $\pi$--decay process 
occurs at the microscopic level. So the $^{3}P_{0}$ QPCM 
largely overestimates majority of the decay rates. 
Discrepancies were found to have an origin in relativistic effects. Starting
from this observation, the QPCM was modified by incorporating those 
corrections which stem from general principles, making it closer to the EEM.
The quality of the results we thus obtain seems to support our prescription.
	Despite large differences in the short distance description of baryons
used in our calculations, we did not find any reason to discriminate them by 
looking at the $\pi$ decay properties once the experimental 
kinematics is considered. 
Introducing a three body force in the description of baryons makes the 
agreement with experiment slightly better in some cases, slightly worse in
others. Probably, the observable we examined is not appropriate to extract any
correlation between the predictions for the spectrum and the decay widths. 
Anyhow as a general conclusion it can be established that 
a unified description of the spectrum and the strong baryonic
decays in a 'non--relativistic' scheme seems plausible. 

\vspace{1cm}

	This work has been partially supported by CICYT under grant 
AEN93-0234 and by DGICYT under grant PB91-0119-C02-01. F.C. acknowledges the
Ministerio de Educaci\'on y Ciencia for a FPI fellowship. B. D. has been
partially supported by Conselleria d'Educaci\'o y Ci\`encia of the 
Generalitat Valenciana.
 
\renewcommand{\thesection}{\Alph{section}}
\renewcommand{\theequation}{\thesection.\arabic{equation}}

\section*{Appendix A}
\setcounter{section}{1}
\setcounter{equation}{0}

	All we can know about the baryonic structure is contained in the
intrinsic wave function $\Psi(\vec{\xi}_{1},\vec{\xi}_{2})$ depending on
the six independents coordinates $\vec{\xi}_{1},\vec{\xi}_{2}$. Alternatively
one can define hyperspherical coordinates in the dimension 6--hyperspace as
an hyperradius $\xi$ defined through

\begin{equation}
\xi^{2} = \vec{\xi}_{1}^{\; 2} + \vec{\xi}_{2}^{\; 2}
\end{equation}

\noindent
an a set $\Omega$ of 5 angles: $\hat{\xi}_{1}$ (the spherical angles of 
$\vec{\xi}_{1}$), $\hat{\xi}_{2}$ (the spherical angles of 
$\vec{\xi}_{2}$), and $\phi$ given by 

\begin{equation}
\begin{array}{ccc}
\xi_{1} & = & \xi \sin \phi \nonumber \\
\xi_{2} & = & \xi \cos \phi \nonumber 
\end{array} 
\end{equation}

\noindent
such that $d\vec{\xi}_{1} \, d\vec{\xi}_{2} = d\xi \, \xi^{5} \, d\Omega =
d\xi \, \xi^{5} \, d\phi \, \sin^{2} \!\phi \cos^{2} \! \phi d\hat{\xi}_{1} \, d\hat{\xi}_{2} $

	In terms of $\Omega$ one can construct a complete set of basis functions
on the hypersphere unity. There are the hyperspherical harmonics (HH)
$\mbox{\sl Y}_{[K]}(\Omega)$, characterized by 5 quantum numbers denoted 
by $[K] \equiv [K, l_{1}, m_{1}, l_{2}, m_{2}]$  \cite{BALLOT}. 

	Then the spatial part of the intrinsic wave function can be expanded 
as:
\begin{eqnarray}
\Psi(\vec{\xi}_{1},\vec{\xi}_{2}) \equiv \Psi (\xi, \Omega) & = & 
\sum_{[K]} \Psi_{[K]}(\xi) \mbox {\sl Y}_{[K]}(\Omega)
\end{eqnarray}

\noindent
satisfying the Schr\"odinger equation:

\begin{equation}
(H_{\mbox {\tiny int}} - E ) \Psi (\xi, \Omega) = 0
\end{equation}

\noindent
where $H_{\mbox {\tiny int}}$ stands for the sum of the internal kinetic 
and the potential energy. 

	Explicitly by defining reduced radial wave functions as

\begin{equation}
\psi_{[K_{i}]} (\xi) = \xi^{5/2} \Psi_{[K_{i}]}(\xi) 
\end{equation}

\noindent
and defining the wave function vector

\begin{equation}
\psi = 
\left( 
\begin{array}{c}
\psi_{[K_{1}]} \\ \vdots \\ \psi_{[K_{n}]}
\end{array}
\right)
\end{equation}

\noindent
one has the set of coupled Schr\"odinger equations:

\begin{equation}
\left\{\frac{1}{2m} \left( \frac{\partial^{2}}{\partial \xi^{2}} -
\frac{\nu_{K}}{\xi^{2}} \right) - V + E \right\} \psi(\xi) = 0
\label{c1}
\end{equation}

\noindent
where $V$ is the matrix of the potential

\begin{equation}
 V_{ij}  =  \langle  Y_{[K_{i}]}(\Omega) | V(\xi, \Omega) |
	 Y_{[K_{j}]}(\Omega) \rangle  
\end{equation}

and $\nu_{K}$ is the diagonal matrix

\begin{equation}
(\nu_{K})_{ij}  =  \left[ (K_{i} + 2 )^{2} - \frac{1}{2} \right] 
\delta_{ij} 
\end{equation}

	The set of eqs. (\ref{c1}) can be solved numerically by Numerov 
integration, turning out to be convenient the introduction of the dimensionless
variable:

\begin{equation}
x = \frac{\sqrt{2} m \xi}{\hbar c}
\end{equation}

\noindent
in terms of which we can finally write

\begin{equation}
\label{d1}
\left\{ - \frac{d^{2}}{dx^{2}} - \frac{\nu_{K}}{x^{2}} 
+ \frac{V(x)}{m} - \frac{E}{m}\right\} \chi(x) = 0
\end{equation}

	Let us note that under a variation of the quark mass 
$m \rightarrow m'$ ($x \rightarrow x'=\frac{m'}{m} x$) (\ref{d1}) transforms
into:

\begin{equation}
\left\{ - \frac{d^{2}}{dx'^{2}} - \frac{\nu_{K}}{x'^{2}} 
+ \frac{\bar{V}(x')}{m'} -  \frac{E'}{m'}\right\} \tilde{\chi}(x') = 0
\end{equation}

\noindent
where
 
\begin{eqnarray}
\bar{V}(x') & \equiv &  \frac{m}{m'}V(x) \label{e8} \\
E' & \equiv &  \frac{m}{m'} E  \\
\tilde{\chi}(x') & \equiv &  \chi(x) 
\end{eqnarray}
 
	So, by redefining the parameters of the potential to satisfy (\ref{e8})
we can obtain an energy spectrum rescaled by a factor $\frac{m}{m'}$ without 
any change in the wave functions.
 
	For physical purposes it is more 
convenient to use linear combinations of HH
with total orbital angular momentum $L$ and definite symmetry, 
$\mbox {\sl Y}_{[K, \mbox {\tiny symmetry}]}^{(L,M)}(\Omega)$ in terms of
which the total intrinsic wave function reads:

\begin{eqnarray}
\left[\Psi_{B}(\vec{\xi}_{1}, \vec{\xi}_{2}) \Phi_{B}(S, M_{S};I,M_{I})
\right]_{JJ_{z}}
& = &  
\sum_{i} \Psi_{B_{i}} (\xi) \left[\sum_{[\mbox {\tiny symmetry}]}
(L_{i} S_{i} J| M_{i} M_{S_{i}} J_{z})
\right. \cdot \nonumber \\
& & \hspace{-1cm}\left. \mbox {\sl Y}_{[K_{i}, \mbox {\tiny symmetry}]}^{(L_{i},M_{i})}
\Phi_{[\mbox {\tiny symmetry}]}(S_{i}, M_{s_{i}}; I_{i}, M_{I_{i}})
\right]
\end{eqnarray}

\noindent
having to be totally symmetric as required by the symmetrization postulate.

\newpage


\begin{center}
{\bf Table captions}
\end{center}

\begin{description}

\item{\bf Table 1.} Fitted values of the parameters of the two--body and 
three--body potentials.

\item{\bf Table 2.} Magnetic moments (in nuclear magnetons) and radii (if 
fm$^{2}$) of the nucleon.

\item{\bf Table 3.} Decay widths (in MeV)  with an 
EEM model. Experimental data from \protect{\cite{PDG94}}.

\item{\bf Table 4.} Values of $\gamma$ obtained from the $NN\pi$ vertex. 

\item{\bf Table 5.} Decay widths (in MeV) with a 
$^{3}P_{0}$ model. Pion wave function as given by 
eq. (\protect\ref{pgaussian}) and $\gamma$ taken from table 4.

\item{\bf Table 6.} Decay widths (in MeV)  with a 
modified $^{3}P_{0}$ QPCM. Pion wave 
function given by eq. (\protect\ref{pgaussian}).

\end{description}

\newpage


\begin{center}
{\bf Figure captions}
\end{center}

\begin{description}

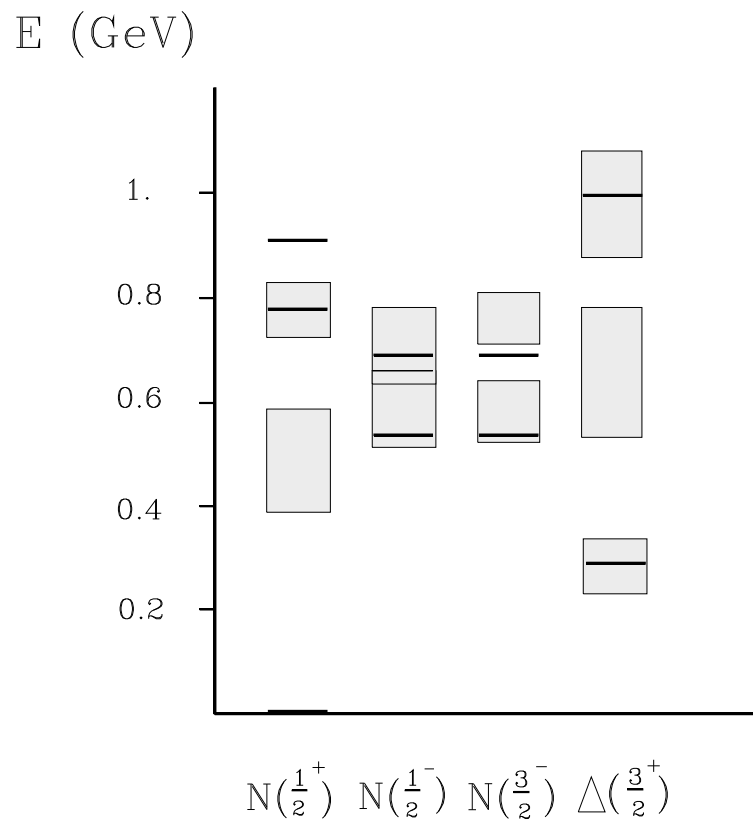
\item{\bf Figure 1.} Relative energy spectrum for positive parity nucleon and delta
and negative parity nucleon states. Solid lines correspond to the predictions
of $V_{I}$ and the shaded region whose size represents the experimental width
to the experimental data \protect\cite{PDG94}.

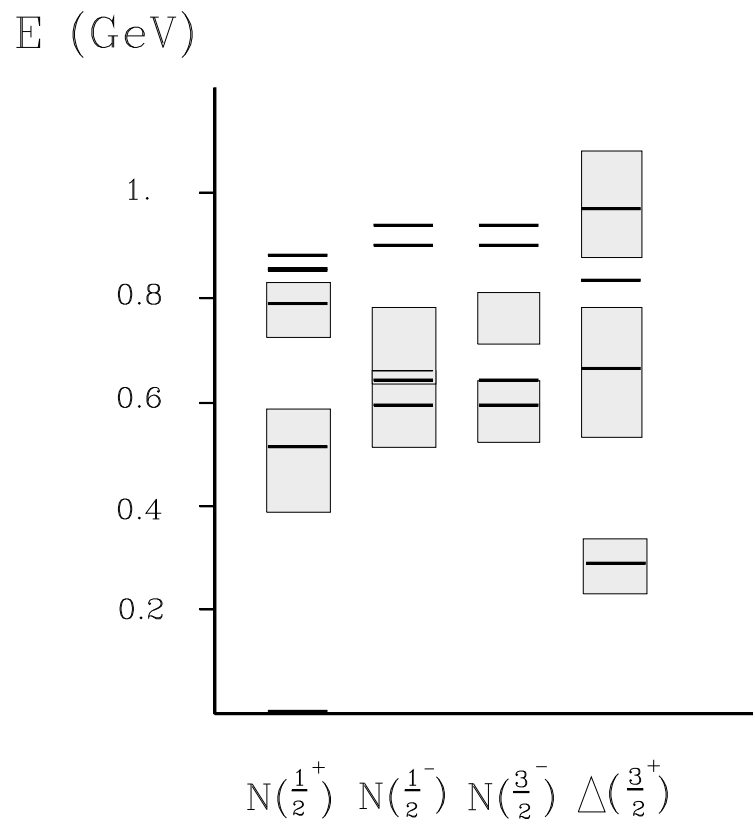
\item{\bf Figure 2.} Idem fig. 1 with $V_{I} \rightarrow V_{II}$. 

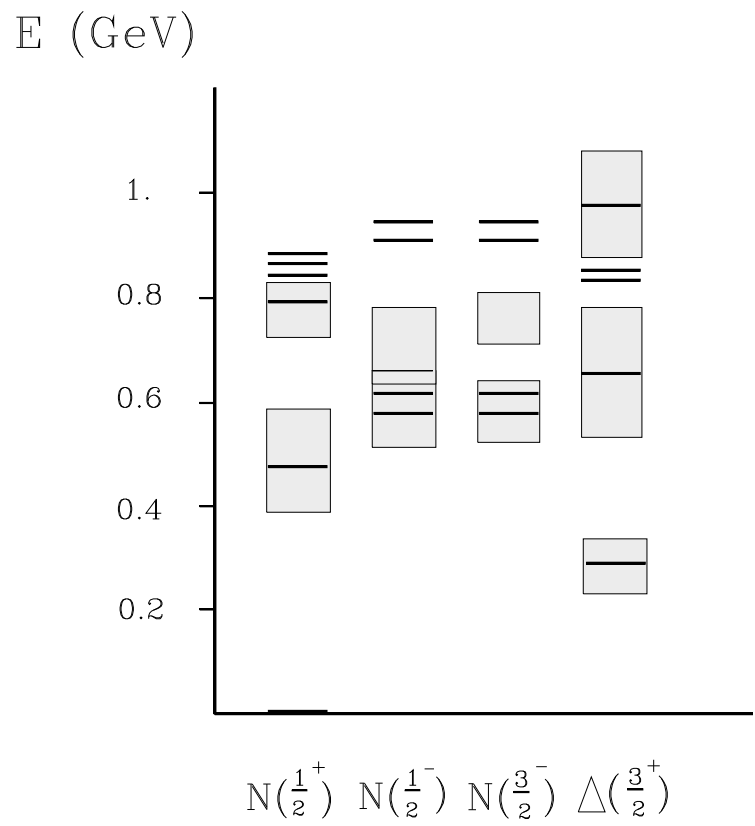
\item{\bf Figure 3.} Idem fig. 1 with $V_{I} \rightarrow V_{III}$. 

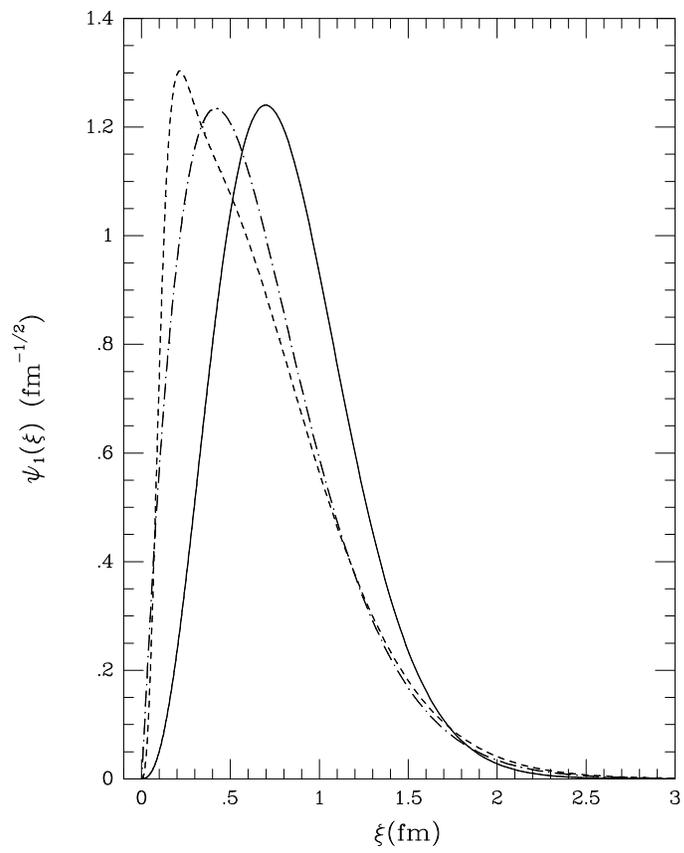
\item{\bf Figure 4.} Spatial symmetric component of the nucleon reduced wave function 
$\psi_{1}(\xi)$. The solid line corresponds to $V_{I}$, the dashed--dot line to
$V_{II}$ and the dashed line to $V_{III}$.

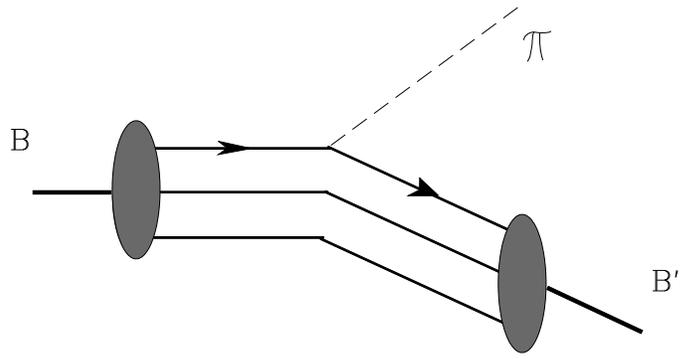
\item{\bf Figure 5.} A baryon B decays through pion emission by 
one of its quarks.

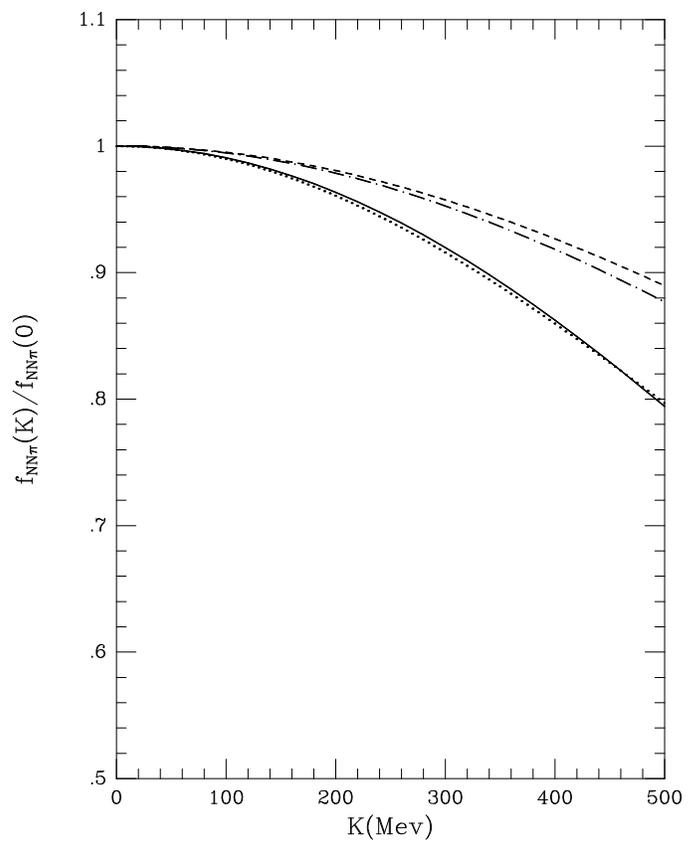
\item{\bf Figure 6.} Normalized $NN\pi$ form factor. The dot line corresponds to the 
phenomenological parameterization of ref. \protect\cite{OSET83}. The solid line,
dashed--dot and dashed line correspond to the predictions of $V_{I}$, $V_{II}$ 
and $V_{III}$ respectively. 

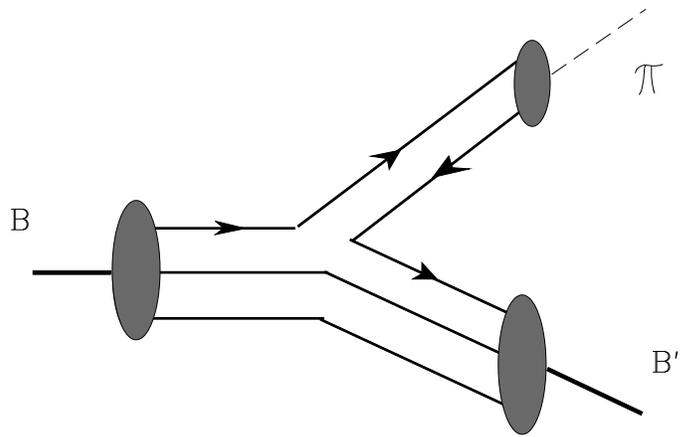
\item{\bf Figure 7.} A quark and antiquark from the vacuum recombine with the 
quarks of the initial baryon to give rise to the final baryon and pion.

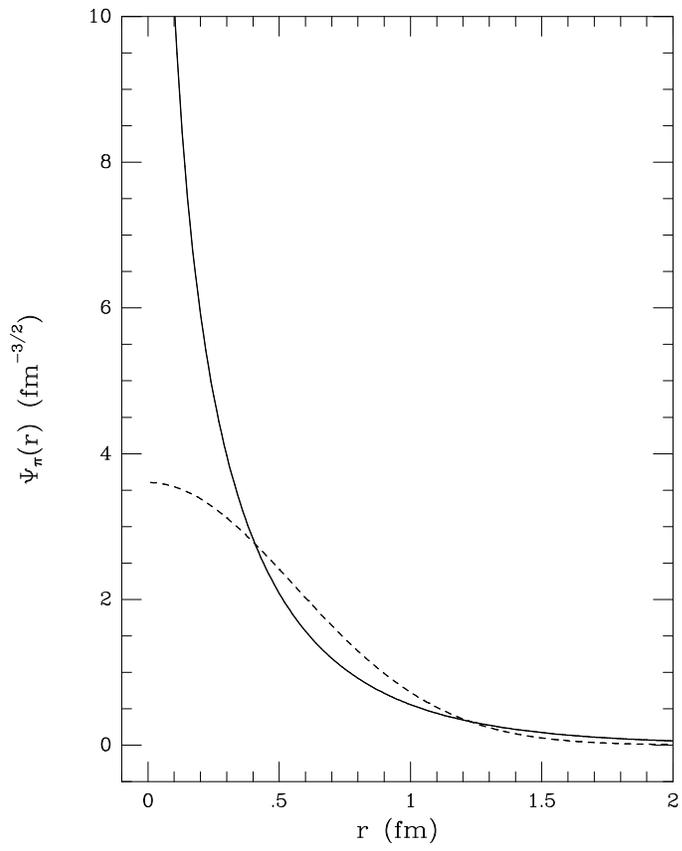
\item{\bf Figure 8.} Normalized pion wave function. The solid line is obtained from the 
Fourier transform of the pion form factor (\protect\ref{electroff}) The dashed
line corresponds to the gaussian form eq. (\protect\ref{pgaussian}).

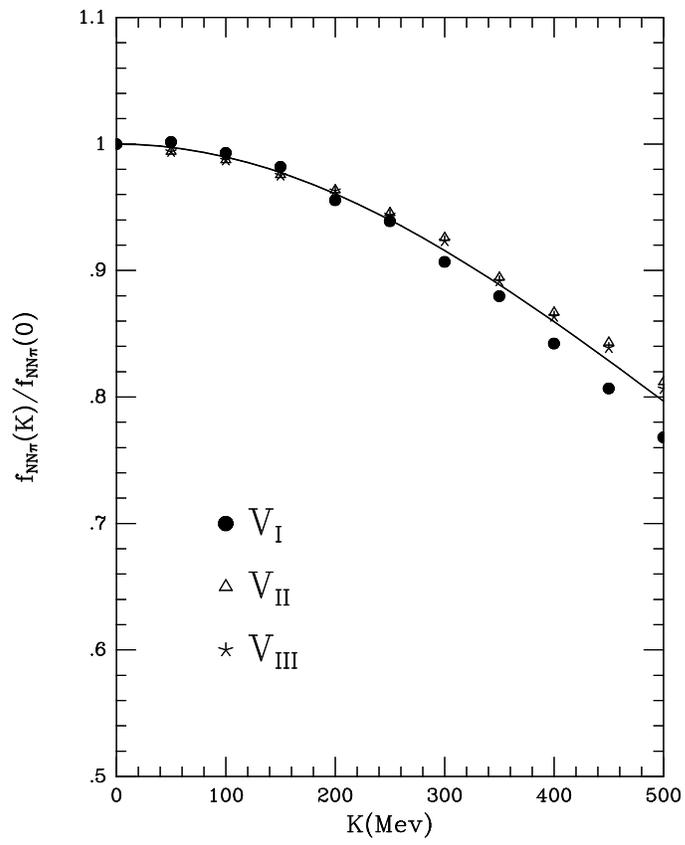
\item{\bf Figure 9.} Normalized $NN\pi$ form factor in the $^{3}P_{0}$ model. 
The line corresponds to the phenomenological parameterization of 
ref. \protect\cite{OSET83}. For the pion
wave function, eq. (\protect\ref{pgaussian}) with $R_{A}^{2}= 8$ GeV$^{-2}$.

\end{description}


\setcounter{table}{10}
\
\vfill
\begin{table}[p]
\begin{center}
\begin{tabular}{|c|c|c|c|c|}
\hline \hline 
\multicolumn{2}{|c|}{\rule{0pt}{3.5ex}}
& $V_{I}$ & $V_{II}$ & $V_{III}$ \\[2ex]
\hline
\multicolumn{2}{|c|}{\rule{0pt}{3.5ex}$m_{u}=m_{d}$ (GeV)} 
& 0.337 & 0.355 & 0.320 \\[2.5ex]
\hline
\rule{0pt}{3.5ex}
$V^{\mbox {\scriptsize (COUL)}}$ & $\kappa$ (GeV fm) 
& 0.1027 & 0.289 & 0.321 \\[2.5ex]
\hline
$V^{(\vec{\sigma} \vec{\sigma})}$ & 
\rule{0pt}{4.5ex} 
\begin{tabular}{l} $\kappa_{\sigma}$ (GeV fm) \\ $r_{0}$ (fm) 
\end{tabular} &
\begin{tabular}{c} 0.1027 \\ 0.4545 
\end{tabular} &
\begin{tabular}{c} 0.049 \\ 0.40
\end{tabular} &
\begin{tabular}{c} 0.044 \\ 0.49
\end{tabular} \\[2.5ex]
\hline
\rule{0pt}{3.5ex}
$V^{\mbox {\scriptsize (CONF)}}$ & $a^{2}$ (GeV$^{-1}$ fm) 
& 1.063 & 4.570 & 4.124 \\[2.5ex]
\hline
\rule{0pt}{4.5ex}
$V_{I}^{(3)}$ & 
\begin{tabular}{l} $V_{0}$ (GeV$^{-2}$ fm$^{-6}$) \\ $m_{0}$ (GeV) 
\end{tabular} &
\begin{tabular}{c} -- \\ --
\end{tabular} &
\begin{tabular}{c} -61.63 \\ 0.25
\end{tabular} &
\begin{tabular}{c} -- \\ --
\end{tabular} \\[2.5ex]
\hline
\rule{0pt}{4.5ex}
$V_{II}^{(3)}$ & 
\begin{tabular}{l} $V_{0}$ (GeV) \\ $\lambda$ (fm) 
\end{tabular} &
\begin{tabular}{c} -- \\ --
\end{tabular} &
\begin{tabular}{c} -- \\ --
\end{tabular} &
\begin{tabular}{c} -35.5 \\ 0.25
\end{tabular} \\[2.5ex]
\hline\hline
\end{tabular}
\vskip 2cm
{\bf \large Table 1}
\label{t1}
\end{center}
\end{table}
\vfill

\addtocounter{table}{1}%

\
\vfill
\begin{table}[p]
\begin{center}
\begin{tabular}{ccccc@{}}
\hline \hline
\rule{0pt}{3.5ex}
& $V_{I}$ & $V_{II}$ & $V_{III}$ & Exp.\\[2ex]
\hline
\rule{0pt}{4.5ex}
$\mu_{p}$ & 2.76 & 2.64 & 2.93 & 2.79 \\ [2.ex]
\rule{0pt}{2.5ex}
$\mu_{n}$ & -1.83 & -1.76 & -1.95 & -1.91 \\ [2.ex]
\rule{0pt}{2.5ex}
$\langle r^{2}_{N} \rangle_{\mbox {\tiny m}}$ & 0.218 & 0.128 & 0.115 &  \\ [2.ex]
\rule{0pt}{2.5ex}
$\langle r^{2}_{p} \rangle_{\mbox {\tiny Ch}}$ & 0.238 & 0.133 & 0.112 & 0.74 $\pm$  0.02 \\ [2.ex]
\rule{0pt}{2.5ex}
$\langle r^{2}_{n} \rangle_{\mbox {\tiny Ch}}$ & -0.02 & -0.005 & -0.004 & -0.119 $\pm$ 0.004\\ [2.5ex]
\hline\hline	
\end{tabular}
\label{t2}
\vskip 2cm
{\bf \large Table 2}
\addtocounter{table}{1}
\end{center}
\end{table}
\vfill

\
\vfill
\begin{table}
\begin{center}
\begin{tabular}{@{}lcccc}
\hline \hline
\rule{0pt}{3.5ex}
& $V_{I}$ & $V_{II}$ & $V_{III}$ & Exp.  \\[2ex]
\hline \hline
\rule{0pt}{4.5ex}
$\Delta(1232) \rightarrow N \pi$ & 
79.6 & 72.1 & 67.7 & 115--125 \\ [2.ex]
\rule{0pt}{2.5ex}
$N(1440) \rightarrow N \pi$ & 
 3.4 & 0.17 & 0.01 & 210--245 \\ [2.ex]
\rule{0pt}{2.5ex}
$N(1440) \rightarrow \Delta \pi$ & 
7.1 & 17.6 & 24.1 & 70--105 \\ [2.ex]
\rule{0pt}{2.5ex}
$\Delta(1600) \rightarrow N \pi$ & 
20.1 & 94.1 & 148.6 & 35--88 \\ [2.ex]
\rule{0pt}{2.5ex}
$\Delta(1600) \rightarrow \Delta \pi$ & 
2.85 & 0.10 & 0.08 & 140--245 \\ [2.ex]
\rule{0pt}{2.5ex}
$N(1520) \rightarrow N \pi$ & 
61.8 & 22.3 & 17.8 & 60--72  \\ [2.ex]
\rule{0pt}{2.5ex}
$N(1520) \rightarrow \Delta \pi$ & 
78.0 & 56.1 & 55.3 & 18--30 \\ [2.ex]
\rule{0pt}{2.5ex}
$N(1535) \rightarrow N \pi$ & 
240 & 149 & 117 & 53--83 \\ [2.ex]
\rule{0pt}{2.5ex}
$N(1535) \rightarrow \Delta \pi$ & 
9.7 & 8.3 & 9.1 & $< $1.5 \\ [2.ex]
\hline\hline	
\end{tabular}
\label{t3}
\vskip 2cm
{\bf \large Table 3}
\addtocounter{table}{1}
\end{center}
\end{table}
\vfill


\
\vfill
\begin{table}[p]
\begin{center}
\begin{tabular}{@{}lccc@{}}
\hline \hline
 \rule{0pt}{4.5ex} & $V_{I}$ & $V_{II}$ & $V_{III}$ \\[2ex]
\hline
\rule{0pt}{3.5ex}
$\gamma$ & 7.02 & 9.78 & 11.04 \\ [1.5ex]
\hline\hline	
\end{tabular}
\label{t4}
\vskip 2cm
{\bf \large Table 4}
\addtocounter{table}{1}
\end{center}
\end{table}


\
\vfill
\begin{table}
\begin{center}
\begin{tabular}{@{}lcccc}
\hline \hline
\rule{0pt}{3.5ex}
& $V_{I}$ & $V_{II}$ & $V_{III}$ & Exp.  \\[2ex]
\hline \hline
\rule{0pt}{4.5ex}
$\Delta(1232) \rightarrow N \pi$ & 
167 & 210 & 241 & 115--125 \\ [2.ex]
\rule{0pt}{2.5ex}
$N(1440) \rightarrow N \pi$ & 
 452 & 1076 & 1509 & 210--245 \\ [2.ex]
\rule{0pt}{2.5ex}
$N(1440) \rightarrow \Delta \pi$ & 
66.5 & 228 & 337 & 70--105 \\ [2.ex]
\rule{0pt}{2.5ex}
$\Delta(1600) \rightarrow N \pi$ & 
19.8 & 0.53 & 7.0 & 35--88 \\ [2.ex]
\rule{0pt}{2.5ex}
$\Delta(1600) \rightarrow \Delta \pi$ & 
255 & 498 & 656 & 140--245 \\ [2.ex]
\rule{0pt}{2.5ex}
$N(1520) \rightarrow N \pi$ & 
268 & 319 & 353 & 60--72  \\ [2.ex]
\rule{0pt}{2.5ex}
$N(1520) \rightarrow \Delta \pi$ & 
532 & 999 & 1203 & 18--30 \\ [2.ex]
\rule{0pt}{2.5ex}
$N(1535) \rightarrow N \pi$ & 
429 & 464 & 410 & 53--83 \\ [2.ex]
\rule{0pt}{2.5ex}
$N(1535) \rightarrow \Delta \pi$ & 
28.1 & 74.0 & 107 & $< $1.5 \\ [2.ex]
\hline\hline	
\end{tabular}
\label{t5}
\vskip 2cm
{\bf \large Table 5}
\addtocounter{table}{1}
\end{center}
\end{table}


\
\vfill
\begin{table}
\begin{center}
\begin{tabular}{@{}lcccc}
\hline \hline
\rule{0pt}{3.5ex}
& $V_{I}$ & $V_{II}$ & $V_{III}$ & Exp.  \\[2ex]
\hline \hline
\rule{0pt}{4.5ex}
$\Delta(1232) \rightarrow N \pi$ & 
88.6 & 112 & 123 & 115--125 \\ [2.ex]
\rule{0pt}{2.5ex}
$N(1440) \rightarrow N \pi$ & 
 114 & 307 & 469 & 210--245 \\ [2.ex]
\rule{0pt}{2.5ex}
$N(1440) \rightarrow \Delta \pi$ & 
27.6 & 116 & 183 & 70--105 \\ [2.ex]
\rule{0pt}{2.5ex}
$\Delta(1600) \rightarrow N \pi$ & 
2.1 & 2.5 & 5.6 & 35--88 \\ [2.ex]
\rule{0pt}{2.5ex}
$\Delta(1600) \rightarrow \Delta \pi$ & 
62.0 & 121 & 177 & 140--245 \\ [2.ex]
\rule{0pt}{2.5ex}
$N(1520) \rightarrow N \pi$ & 
95.1 & 105 & 118 & 60--72  \\ [2.ex]
\rule{0pt}{2.5ex}
$N(1520) \rightarrow \Delta \pi$ & 
45.9 & 75.2 & 111 & 18--30 \\ [2.ex]
\rule{0pt}{2.5ex}
$N(1535) \rightarrow N \pi$ & 
49.2 & 44.1 & 55.8 & 53--83 \\ [2.ex]
\rule{0pt}{2.5ex}
$N(1535) \rightarrow \Delta \pi$ & 
15.3 & 36.3 & 52.1 & $< $1.5 \\ [2.ex]
\hline\hline	
\end{tabular}
\label{t6}
\vskip 2cm
{\bf \large Table 6}
\end{center}
\end{table}

\end{document}